\documentstyle[12pt]{article}
\begin{document}
{\hskip 11cm}
BIHEP-TH-95-11
\par
{\hskip 11cm}
PUTP-95-09\\
\par
\begin{large}
\begin{center}
\begin{bf}
Glueball Interpretation of $\xi$(2230)\\
\end{bf}
\end{center}
\end{large}
\begin{center}
Tao Huang, Shan Jin, Da-hua Zhang\\
\begin{it}
        CCAST ( World Laboratory ), P.O. Box 8730, Beijing 100080, and\\
        Institute of High Energy Physics,
        P.O.Box 918, Beijing 100039, China\\
\end{it}
\end{center}
\begin{center}
        Kuang-Ta Chao\\
\begin{it}
        CCAST ( World Laboratory ) and\\
        Department of Physics, Peking University, Beijing 100871, China\\
\end{it}
\end{center}
%\date{}
%\maketitle
\par
\begin{small}
\begin{abstract}
{\hskip 0cm}
            On the basis of the results of $\xi(2230)\rightarrow\pi^{+}\pi^{-},
            p\bar{p}$ and $K\bar{K}$, measured by the BES Collaboration in
            radiative J/$\psi$ decays, combined with the upper limit of
            Br($\xi\rightarrow p\bar{p}$)Br($\xi\rightarrow K\bar{K}$),
            measured by PS185 experiment,
            we argue that
            the distinctive properties of $\xi$(2230), the flavor-symmetric
            decays and the narrow partial decay widths to $\pi\pi$ and
            $K\bar{K}$ as well as its
            copious production in radiative  J/$\psi$ decay,
            would strongly favor the glueball interpretation of $\xi$(2230).\\
\par
            PACS number(s): 14.40.Cs, 12.38.Aw, 11.30.Hv, 13.20.Gd
\end{abstract}
\par
\begin{bf}
{\hskip -0.8cm}
\end{bf}
\par
{\hskip 0cm}
            BES Collaboration have reported their new results about $\xi$(2230)
            $^{[1]}$. They confirmed the existence of $\xi$(2230) in the
            decays J/$\psi \rightarrow \gamma K\bar{K}$ and they also found
            two new non-strange decay modes
            $\xi \rightarrow p\bar{p}$ and $\pi^{+} \pi^{-}$.
            The preliminary results of the mass, width and branching ratios
            of $\xi$(2230) measured by BES are the following$^{[1]}$:
            In J/$\psi \rightarrow \gamma \pi^{+}\pi^{-}$ process,
            M$_{\xi}$ = (2235$^{+}_{-}4^{+}_{-}$6)MeV,
            $\Gamma_{\xi}$ = (19$^{+13 +}_{-11 -}$12)MeV,
            Br$(J/\psi\rightarrow\gamma \xi)$ Br$(\xi\rightarrow\pi^{+}\pi
            ^{-}) = (5.6^{+1.8 +}_{-1.6 -}1.4)\times 10^{-5}$.
            In J/$\psi \rightarrow \gamma K^{+} K^{-}$ process,
            M$_{\xi}$ = (2230$^{+6 +}_{-7 -}$12)MeV,
            $\Gamma_{\xi}$ = (20$^{+20 +}_{-15 -}$12)MeV,
            Br$(J/\psi\rightarrow \gamma \xi)$ Br$(\xi\rightarrow K^{+} K
            ^{-}) = (3.3^{+1.6 +}_{-1.3 -} 1.1)\times 10^{-5}$.
            In J/$\psi \rightarrow \gamma p\bar{p}$ process,
            M$_{\xi}$ = (2235$^{+}_{-} 4 ^{+}_{-}$5)MeV,
            $\Gamma_{\xi}$ = (15$^{+12 +}_{-9 -}$9)MeV,
            Br$(J/\psi\rightarrow\gamma \xi)$Br$(\xi\rightarrow p\bar{p}
            ) = (1.5^{+0.6 +}_{-0.5 -} 0.5)\times 10^{-5}$.\\
\par
{\hskip 0cm}
            Previous theoretical interpretations of $\xi$(2230)
            included
            identification as a high spin $s\bar{s}$ state$^{[2]}$, a
            multiquark state (such as a 4-quark state$^{[3,4]}$,
            a $\Lambda\bar{\Lambda}$ bound
            state$^{[5]}$,
            a neutral color scalar bound state $^{[6]}$,
            etc.), a hybrid
            state$^{[3,7]}$ and a glueball$^{[8]}$.
            Since Mark III Collaboration only found strange decay modes of
            $\xi$(2230)$^{[9]}$, the
            $s\bar{s}$ interpretation seemed plausible and the glueball
            interpretation would be ruled out. The new discovery of BES
            experiments on non-strange decay modes  provides us with extremly
            important clue so that the nature of $\xi$(2230) will
            become clear.\\
\par
{\hskip 0cm}
            Compared with other mesons, $\xi(2230)$ has many distinctive
            properties:
            (1) flavor-symmtric decays.
            With the phase space factors removed, it can be found that
            the probability
            of $\xi \rightarrow \pi^{+}\pi^{-}$ is of the same order
            as that of
            $\xi \rightarrow K^{+}K^{-}$.
            (2) Copiously production in radiative J/$\psi$ decays$^{[10]}$.
            From the upper limit$^{[11,12]}$ Br$(\xi \rightarrow p\bar{p})$
            Br$(\xi \rightarrow K\bar{K}) < 1 \times 10^{-4}$,
            where $K\bar{K}$ include all kaon pairs,
            and the BES's results, one can easily estimated that lower
            limit of Br$(J/\psi \rightarrow \gamma \xi$) is
            $3 \times 10^{-3}$.
            (3) Narrow width.
            Both Mark III's results and BES's results show that the width
            of $\xi$(2230) is only about 20 MeV$^{[1,9]}$. In this paper,
            we use $\Gamma_{\xi}$= 20 MeV.
            From (2)(3), it can also be easily estimated that
            Br$(\xi \rightarrow K^{+} K^{-})$ and Br($\xi \rightarrow \pi^{+}
            \pi^{-})$ are smaller than 2$\%$, so the partial width
            $\Gamma_{\pi^{+}\pi^{-}}$ and
            $\Gamma_{K^{+}K^{-}}$ are smaller than 400 keV$^{[10]}$.\\
\par
{\hskip 0cm}
            Although the knowledge about $\xi$(2230) is still limited by
            the experimental facts,
            the features of $\xi$(2230) we have
            known so far are so special and so clear that they would make it
            possible for us to understand the nature of $\xi$(2230).\\
\par
{\hskip 0cm}
            For an $s\bar{s}$ meson, it should
            not show so good flavor-symmetric decay behavior and the
            $\Gamma_{K^{+}K^{-}}$ would not be so narrow since its
            decay to $K\bar{K}$ is OZI allowed. The
            resonance aroud 2.2 GeV found by LASS Collaboration$^{[13]}$
            in $Kp \rightarrow \Lambda K\bar{K}$ process may be different
            from $\xi$(2230) since it is produced in different mechanism
            from the radiative J/$\psi$ decay and its mass, width
            and J$^{PC}$ are
            different from the BES's results$^{[1,14]}$.
            More generally, considering all conventional $q\bar{q}$ mesons,
            including $(u\bar{u}+d\bar{d})$, $s\bar{s}$ or their admixtures,
            it is worth noticing that there are
            not any other particles showing such properties$^{[12]}$ as
            $\xi$ except for
            the particles with pure OZI suppressed decay modes such as
$J/\psi$,
            $\chi_{c0},\chi_{c2}$, etc.
            The typical width for conventional
            $q\bar{q}$ mesons with OZI allowed decay modes is about
            100$-$200 MeV
            if all other quantum numbers are allowed and the phase space
            is not too small, especially its partial width of certain main
            decay modes must be of order 10-100 MeV. E.g., for the $P-$wave
            $J^{PC}$=$2^{++}$ mesons, the $f_2(1270)$, which is mainly a
            $(u\bar{u}+d\bar{d})$ state, has a partial width  of about 150 MeV
            to $\pi\pi$, while the $f'_2(1525)$, which is mainly an $s\bar{s}$
            state, has a partial decay width of about 50 MeV to $K\bar{K}$
            $^{[12]}$. For the $F-$wave mesons, based on some quark model
            calculation it was argued$^{[2]}$ that if $\xi(2230)$ were an
            $^3F_2$ or $^3F_4$ $s\bar{s}$ state, its decay to $K\bar{K}$ could
            be suppressed by the L = 3 centrifugal barrier and consequently
            the decay width to $K\bar{K}$ could be lowered to 20$-$30 MeV, but
            cannot be as small as the order of several hundreds keV.
            On the other hand,
            the $f_4(2050)$, which is a $(u\bar{u}+d\bar{d})$ dominant $^3F_4$
            state, has an observed total width of 200 MeV and a partial decay
            width of 30 MeV to $\pi\pi$$^{[12]}$. We see that, with both
            experimental observations and quark model calculations, all this
            kind of $q\bar{q}$ states can hardly have a total width of 20 MeV
            and, in particular, cannot have a partial decay width of the order
            of several hundreds keV to $\pi\pi$ or $K\bar{K}$, as observed
            for $\xi(2230)$. Therefore, as a result of the observation of
            small partial widths of $\xi\rightarrow\pi\pi, K\bar{K}$,
            we may conclude that the $\xi(2230)$ cannot be a conventional
            $q\bar{q}$ meson.\\
\par
{\hskip 0cm}
            The copiously production in radiative $J/\psi$ decay would
            disfavor the interpretation of a multiquark state such as
            a $\Lambda\bar{\Lambda}$
            bound state, a 4-quark state, etc.
            The production rate of $\xi(2230)$ could be only
            smaller than $\eta_c$ and $\eta'(958)$, but larger than or as
            large as $\iota(1440)$,$\theta(1710)$,$f_4(2050)$,$f_2(1270)$ and
            $f_2'(1525)$. Thus $\xi(2230)$ is even more copiously produced
            than some well established conventional $q\bar{q}$ mesons such as
            $f_2(1270)$ and $f_2'(1525)$. As for multiquark states,
            according to the naive quark pair counting rule, they are
            usually expected to have smaller production rates than the
            corresponding $q\bar{q}$ states, since the creation of more quark
            pairs is needed for multiquark state production. Most naturally,
            the rich production of $\xi$ in radiative $J/\psi$ decays will
            imply that the $\xi(2230)$ is likely to be a glueball or a
            $q\bar{q}g$ hybrid state, but the former should have an even
            larger production rate than the latter.
            As for hybrid interpretation, the width of a hybrid should not be
            so narrow
            since its decay is not totally OZI suppressed (with only one
            gluon converted into a quark pair), thus it would face much trouble
            in explaining the narrow width of $\xi$(2230) ( esp. the partial
            widths $\Gamma_{\pi\pi}$ and $\Gamma_{KK}$ ).\\
\par
{\hskip 0cm}
            Finally, let's consider the glueball interpretation. So far,
            the gluball interpretation has no conflict with all the properties
            of $\xi$(2230). \\
\par
{\hskip 0cm}
            The mesonic decay of glueballs is determined by
            their flavor SU(3) singlet nature.
            With phase space factors removed,
            glueballs are naively expected to couple equally to all flavors.
            Since there has been no glueball
            confirmed by expriments, the best way looking into the flavor
            symmetry  should be to study the decay processes which proceed
            through a two gluon intermediate state$^{[10]}$. Fortunetly,
            there are a lot
            of experimets which have already studied such processes as the
            decays of charmonium family. One example is, the two gluon
            system in radiative J/$\psi$ decays is an SU(3) singlet. This
            predicts$^{[15]}$ the ratio R=$\Gamma(J/\psi\rightarrow\gamma
            f'_2(1525))/\Gamma(J/\psi\rightarrow\gamma f_2(1270))$ = 0.45,
            if phace-space corrections are considered. The experimental result
            R = 0.46$^{+}_{-}$0.07$^{[12]}$ is in good agreement with the
            SU(3) singlet prediction.
            Other examples are that both
            $\chi_{c0}$ and $\chi_{c2}$ show flavor-symmetric decay
            behavior in their mesonic decays$^{[12]}$.
            We believe the observed flavor-symmetry pattern of charmoniun
            decays does lend strong support to the conjecture that the glueball
            decays should be flavor-symmetric.\\
\par
{\hskip 0cm}
            The copious production in radiative J/$\psi$ decay is just
            the expectation for a glueball if we naively count the vertex
            of Feynman Diagram. The  production rate of a glueball is
            of the order $\alpha \alpha_s^{2}$ while the production rate of
            a conventional $q \bar{q}$ meson is of the order $\alpha
            \alpha_s^{4}$. So a glueball could be easier produced than a
            conventional $q \bar{q}$ meson with the same $J^{PC}$.\\
\par
{\hskip 0cm}
            The narrow width is also naively expected by conventional
            understanding of glueballs since their dacay to $q\bar{q}$ state is
            OZI suppressed and the suppression only acts at one vertex
            because of the absence of the initial $q\bar{q}$ annihilation for
            a glueball decay$^{[16]}$ . For example,
            the narrow width of $\xi$(2230) can be easily explained
            by naive estimation that the width
            of a glueball $\Gamma_{G}$ is about
            $\sqrt{\Gamma_
            {f_2 (1270)} \Gamma_{\chi_{c2}}}$, i.e., about
            10-50 MeV.\\
\par
{\hskip 0cm}
            It is not surprising that the branching ratio
            Br$(\xi \rightarrow K^{+} K^{-})$ and Br($\xi \rightarrow \pi^{+}
            \pi^{-})$ are smaller than 2$\%$
            ( consequently,
            $\Gamma_{\pi^{+}\pi^{-}}$ and
            $\Gamma_{K^{+}K^{-}}$ are smaller than 400 keV) for
            a glueball$^{[10]}$.
            From the knowledge $^{[12]}$
            about the hadronic decays of J/$\psi$,$\eta_c$,
            $\chi_{c0}$ and $\chi_{c2}$
            which proceed through pure gluon intermediate state the same as
            glueball decays, it may be
            naively expected that another possible feature of the glueball
            decay is that the glueballs probably have more decay modes than
            conventional $q\bar{q}$ states.
            A $q\bar{q}$ meson decay occurs
            when the color flux tube formed by $q$ and $\bar{q}$ is broken
            at large distances by the creation of new quark pairs ( the OZI
            allowed decay); whereas a glueball decay proceeds via the gluon
            hadronization. There are more possibilities and combinations for
            the gluon fragmentation and hadronization than for the quark pair
            creation in a color flux tube. Therefore, a glueball may have more
            decay modes than a $q\bar{q}$ meson, and hence have smaller
            branching ratios to many final states. In this connections, for
            $\xi(2230)$ the observed flavor-symmetric decays to $\pi\pi$,
            $K\bar{K}$ and the smallness of these decay branching ratios seem
            to favor the assignment that the $\xi(2230)$ is a glueball.\\
\par
{\hskip 0cm}
            Since the observed $\pi\pi,K\bar{K},p\bar{p}$ are expected to be,
            according the above discussion, only a small portion of the decay
            modes of $\xi$, searches for more decay modes of $\xi(2230)$ may be
            important. A systematical test of the flavor-symmetric nature in
            the decays will be meaningful for the glueball interpretation of
            $\xi$.
            We have noticed that the $p\bar{p},
            \pi\pi$ and $K\bar{K}$ decay modes of a particle
            are the easiest
            tagged modes with high efficiency and low background for the BES
            detector.
            Other decay modes, such as $\eta\eta$, $\eta\eta'$,$\eta'\eta'$,
            $\rho\rho$, $K^{*}K^{*}$, $\omega\omega$, $\phi\phi$,
            $\pi\pi\pi\pi$, ${\pi\pi}KK$, etc.,
            may suffer from either too low detecting efficiency or too large
            backgrounds or both of them.
            Anyway, we are interested in the
            results on such decay modes from BES or other experiments.\\
\par
{\hskip 0cm}
            It is interesting to note that a comprehensive lattice study of
            SU(3) glueballs by the UKQCD Collaboration suggests the mass of the
            $2^{++}$ glueball be $2270_{-}^{+}100$ MeV $^{[17]}$. Then in the
            connection to the mass of $\xi(2230)$ and its glueball
            interpretation, it might be suggested that the
            spin-parity of $\xi$ should be $2^{++}$. We hope the BES
            experiments will collect more statistics of $J/\psi$ data to obtain
            more definite result of the spin-parity of $\xi(2230)$ after
            the BES detector upgrade.\\
\par
{\hskip 0cm}
            In summary, we believe that the recent results reported by BES
            is very encouraging in the identification of the puzzling state
            $\xi(2230)$. With both the BES and the PS185 experiments, this
            particle is found to have striking features that it has flavor-
            symmetric couplings to $\pi\pi$ and $K\bar{K}$, a large production
            rate in radiative $J/\psi$ decays and very narrow partial decay
            widths to $\pi\pi$ and $K\bar{K}$.
            The $q\bar{q}$ model, multiquark model and hybrid model
            would face many difficulties in explaining the special
            properties of $\xi$(2230). On the contrary, the glueball
            interpretation can naturally explain all of them. Therefore,
            these features would strongly favor the glueball interpretation
            of $\xi$(2230).\\
\par
{\hskip 0cm}
            This work was supported in part by the National Natural Science
            Foundation of China, and the State Education Commission of China.

\end{small}

\begin{thebibliography}{99}
\bibitem{1}S. Jin, talk given at the International Workshop on Hadron Physics
           at Electron-Positron Colliders, Beijing, October 14-17, 1994;\\
           J. Li talk given at the 27th International Conference on High Energy
           Physics, Glasgow, July 21-27, 1994.
\bibitem{2}S. Godfrey, R. Kokoski, and N. Isgur, Phys. Lett. B 141 (1984) 439.
\bibitem{3}K. T. Chao, Phys. Rev. Lett. 60 (1988) 2579;\\
           K. T. Chao Commun. Theor. Phys. 3(1984)757.
\bibitem{4}S. Pakvasa, M. Suzuki and S.F. Tuan, Phys. Lett. B 145 (1984) 135.
\bibitem{5}S. Ono, Phys. Rev. D35 (1987) 944
\bibitem{6}M.P. Shatz, Phys. Lett. B 138 (1984) 209.
\bibitem{7}M.S. Chanowitz and S. R. Sharpe, Phys. Lett. B 132 (1983) 413;\\
           M. Le Yaouanc et al., Z. Phys. C28(1985)309.
\bibitem{8}B.F.L. Ward, Phys. Rev. D31 (1985) 2849.
\bibitem{9}R.M. Baltrusaitis et al., Phys. Rev. Lett. 56 (1986) 107.
\bibitem{10}K. T. Chao, PUTP-94-26 ( hep-ph/9502408 ), talk given at the
            CCAST Workshop on Tau-Charm Factory, Beijing, November 1994.
\bibitem{11}P.D. Barnes et al., Phys. Lett. B 309 (1993) 469.
\bibitem{12}Particle Data Group, L. Montanet et al., Phys. Rev. D50 3-I (1994)
            1173.
\bibitem{13}D. Aston et al., Phys. Lett. B 215 ( 1988) 199; Nucl. Phys. B301
            (1988) 525.
\bibitem{14}The very prelimilary result of spin-parity analysis from the BES
            experiments is that $J^{PC}=2^{++}$ is prefered to $J^{PC}=4^{++}$
            ( see ref [1] ).
\bibitem{15}H. J. Lipkin and H. R. Rubenstein, Phys. Lett 76B (1978) 324.
\bibitem{16}D. Robson, Nucl. Phys. B130 (1977) 328.
\bibitem{17}G. B. Bali et al., Phys. Lett. B309 (1993) 378.
\end{thebibliography}
\end{document}